\documentclass[fleqn,10pt]{article}

\RequirePackage[utf8]{inputenc}
\RequirePackage[english]{babel}

\RequirePackage{ifthen}
\RequirePackage{calc}
\AtEndOfClass{\RequirePackage{microtype}}
\RequirePackage{times}      
\RequirePackage{mathptmx}   
\RequirePackage{ifpdf}

\RequirePackage{amsmath,amsfonts,amssymb}
\RequirePackage{graphicx,xcolor}
\RequirePackage{booktabs}

\RequirePackage{authblk}
\RequirePackage[left=2cm,%
                right=2cm,%
                top=2.25cm,%
                bottom=2.25cm,%
                headheight=12pt,%
                letterpaper]{geometry}%
                
\RequirePackage[labelfont={bf,sf},%
                labelsep=period,%
                justification=raggedright]{caption}

\RequirePackage[colorlinks=true, allcolors=blue]{hyperref}

\RequirePackage[superscript,biblabel]{cite}
              
\RequirePackage{fancyhdr}  
\RequirePackage{lastpage}  
\RequirePackage[explicit]{titlesec}
\titleformat{\section}
  {\color{color1}\large\sffamily\bfseries}
  {\thesection}
  {0.5em}
  {#1}
  []
\titleformat{name=\section,numberless}
  {\color{color1}\large\sffamily\bfseries}
  {}
  {0em}
  {#1}
  []  
\titleformat{\subsection}
  {\sffamily\bfseries}
  {\thesubsection}
  {0.5em}
  {#1}
  []
\titleformat{\subsubsection}
  {\sffamily\small\bfseries\itshape}
  {\thesubsubsection}
  {0.5em}
  {#1}
  []    
\titleformat{\paragraph}[runin]
  {\sffamily\small\bfseries}
  {}
  {0em}
  {#1} 
\titlespacing*{\section}{0pc}{3ex \@plus4pt \@minus3pt}{5pt}
\titlespacing*{\subsection}{0pc}{2.5ex \@plus3pt \@minus2pt}{0pt}
\titlespacing*{\subsubsection}{0pc}{2ex \@plus2.5pt \@minus1.5pt}{0pt}
\titlespacing*{\paragraph}{0pc}{1.5ex \@plus2pt \@minus1pt}{10pt}

\usepackage{titletoc}
\contentsmargin{0cm}
\titlecontents{section}[\tocsep]
  {\addvspace{4pt}\small\bfseries\sffamily}
  {\contentslabel[\thecontentslabel]{\tocsep}}
  {}
  {\hfill\thecontentspage}
  []
\titlecontents{subsection}[\tocsep]
  {\addvspace{2pt}\small\sffamily}
  {\contentslabel[\thecontentslabel]{\tocsep}}
  {}
  {\ \titlerule*[.5pc]{.}\ \thecontentspage}
  []
\titlecontents*{subsubsection}[\tocsep]
  {\footnotesize\sffamily}
  {}
  {}
  {}
  [\ \textbullet\ ]  
  
\RequirePackage{enumitem}

\renewcommand{\@biblabel}[1]{\bfseries\color{color1}#1.}

\newcommand{\keywords}[1]{\def\@keywords{#1}}

\def\xabstract{abstract}
\long\def\abstract#1\end#2{\def\two{#2}\ifx\two\xabstract 
\long\gdef\theabstract{\ignorespaces#1}
\def\go{\end{abstract}}\else
\typeout{^^J^^J PLEASE DO NOT USE ANY \string\begin\space \string\end^^J
COMMANDS WITHIN ABSTRACT^^J^^J}#1\end{#2}
\gdef\theabstract{\vskip12pt BADLY FORMED ABSTRACT: PLEASE DO
NOT USE {\tt\string\begin...\string\end} COMMANDS WITHIN
THE ABSTRACT\vskip12pt}\let\go\relax\fi
\go}

\renewcommand{\@maketitle}{%
{%
\thispagestyle{empty}%
\vskip-36pt%
{\raggedright\sffamily\bfseries\fontsize{20}{25}\selectfont \@title\par}%
\vskip10pt
{\raggedright\sffamily\fontsize{12}{16}\selectfont  \@author\par}
\vskip18pt%
{%
\noindent
{\parbox{\dimexpr\linewidth-2\fboxsep\relax}{\color{color1}\large\sffamily\textbf{ABSTRACT}}}
}%
\vskip10pt
{%
\noindent
\colorbox{color2}{%
\parbox{\dimexpr\linewidth-2\fboxsep\relax}{%
\sffamily\small\textbf\\\theabstract
}%
}%
}%
\vskip25pt%
}%
}%
\definecolor{color1}{RGB}{0,0,0} 
\definecolor{color2}{gray}{1} 

\newlength{\tocsep} 
\usepackage{lipsum} 
\let\oldbibliography\thebibliography
\renewcommand{\thebibliography}[1]{%
\addcontentsline{toc}{section}{\hspace*{-\tocsep}\refname}%
\oldbibliography{#1}%
\setlength\itemsep{0pt}%
}

\usepackage{braket}
\renewcommand{\vec}[1]{\mathbf{#1}}

\usepackage{pdfpages}

\title{Superbunching and Nonclassicality as new Hallmarks of Superradiance}

\author[1,2,*]{Daniel Bhatti}
\author[3]{Joachim von Zanthier}
\author[2]{Girish S. Agarwal}
\affil[1]{Institut f\"ur Optik, Information und Photonik, Universit\"at Erlangen-N\"urnberg, 91058 Erlangen, Germany}
\affil[2]{Department of Physics, Oklahoma State University, Stillwater, OK, USA}
\affil[3]{Erlangen Graduate School in Advanced Optical Technologies (SAOT), Universit\"at Erlangen-N\"urnberg, 91052 Erlangen, Germany}

\affil[*]{daniel.bhatti@fau.de}

\begin{abstract}
Superradiance, i.e., spontaneous emission of coherent radiation by an ensemble of identical two-level atoms in collective states introduced by Dicke in 1954, is one of the enigmatic problems of quantum optics. The startling gist is that even though the atoms have no dipole moment they radiate with increased intensity in particular directions. 
Following the advances in our understanding of superradiant emission by atoms in entangled $W$ states we examine the quantum statistical properties of superradiance. Such investigations require the system to have at least two excitations as one needs to explore the photon-photon correlations of the radiation emitted by such states. We present specifically results for the spatially resolved photon-photon correlations 
of systems prepared in doubly excited $W$ states and give conditions when the atomic system emits nonclassial light. Equally, we derive the conditions for the occurrence of bunching and even of superbunching, a rare phenomenon otherwise known only from nonclassical states of light like the squeezed vacuum. We finally investigate the photon-photon cross correlations of the spontaneously scattered light and highlight the nonclassicalty of such correlations.
\end{abstract}
\begin{document}

\flushbottom
\maketitle
%
%
\thispagestyle{empty}

\section*{Introduction}

Dicke \cite{Dicke(1954)} predicted that if an ensemble of two-level atoms is prepared in a collective state where half of the atoms are in the excited state and half of the atoms are in the ground state the spontaneous emission 
is proportional to the square of the number of atoms as if the particles would radiate coherently in phase like synchronized antennas \cite{Eberly(1975)}. To analyze the phenomenon Dicke introduced the concept of collective spins where $N$ two level atoms are described by the collective spin eigenstates $\ket{N/2, M}$, with 
$M$ running from $M = -N/2, \ldots, +N/2$ in steps of unity. 
Among these states the state $\ket{N/2, 0}$ radiates with an intensity $N^2$ times as strong as that of a single atom. The origin of superradiance is difficult to see since all states $\ket{N/2, M}$  exhibit no macroscopic dipole moment whereas such a dipole moment is commonly assumed to be required for a radiation rate proportional to $N^2$. The reason is that the Dicke states display strong quantum entanglement. The entangled character of the states is particularly apparent for the case of two two-level atoms where the individual atomic states are labeled by $\ket{e_l}$ and $\ket{g_l}$, $l = 1, 2$, for the excited and ground state, respectively. In this case the Dicke state $\ket{1,0} = 1/\sqrt{2} \left( \ket{e_1, g_2} + \ket{g_1, e_2} \right)$, also known as the Bell state or the EPR state, is clearly maximally entangled. For three atoms one of the Dicke states is denoted by $\ket{3/2,-1/2}$, which in current language would be the $W$ state $1/\sqrt{3} \left( \ket{e_1, g_2, g_3} + \ket{g_1, e_2, g_3} + \ket{g_1, g_2, e_3} \right)$ \cite{Dur(2000)}. The single excited generalized $W$ state, where only one atom is excited and $N-1$ atoms are in the ground state, is also known to be fully entangled and plays a particularly important role for single-photon superradiance \cite{Scully(2006),Scully(2007),Scully(2008),Scully(2009),Scully(2009)Super}. 
In fact, it has been recognized that most of the important aspects of superradiance \cite{Dicke(1954),Eberly(1970),Rehler(1971),AGARWAL(1974),Haroche(1982)} can be studied by examining samples in single excited generalized $W$ states
\cite{Nienhuis(1987),Scully(2006),Scully(2007),Scully(2009)Super,Scully(2009),Svidzinsky(2010),Wiegner(2011)PRA,Kaiser(2012),Kaiser(2013),Rohlsberger(2013)} as the emission from these states possesses all the features of superradiance that originally were calculated for samples with an arbitrary number of excitations \cite{Scully(2007),Wiegner(2011)PRA}. The spatial features of one photon superradiance have been extensively studied for example from the perspective of \textit{timed Dicke states} \cite{Scully(2006),Scully(2007),Scully(2009)Super} and also the spectral and temporal aspects have been investigated in a large variety of systems \cite{Scully(2009),Svidzinsky(2010),Cleland(2010),Simmonds(2011),Nori(2011),Nori(2012),Rohlsberger(2013)}. 
Note that a number of recent works \cite{Thiel(2007),Vitanov(2008),Vitanov(2008)a,Zeilinger(2009),Weinfurter(2009),Wineland(2009),Lemr(2009)} have also discussed how
single excited generalized $W$ states for a small number of atoms can be prepared in the laboratory.

The single excited generalized $W$ state does however not allow one to study the quantum statistical properties of superradiance. In order to explore these aspects the system must emit at least two photons. Only then one has access to the photon-photon correlations which display amongst others the particular quantum characteristics of the spontaneously scattered radiation \cite{Glauber(1963)Quantum,Mandel(1977),WALLS(1994)}. To this end it is required to investigate what we will term two-photon superradiance from generalized $W$ states with $n_{e}\geq2$ excitations. 

In the present paper we show that in two-photon superradiance the emitted radiation can exhibit both bunched as well as nonclassical and antibunched light depending on the angle of observation, i.e., the position of the detectors collecting the scattered photons, and on the particular $W$ state, i.e., the number of atoms $N$ and the number of excitations $n_{e}$, considered. In particular, in certain cases it is also possible to observe the phenomenon of superbunching, i.e., photon-photon correlations larger than those maximally measurable for classical light sources. In all the cases the mean intensity displays the familiar features of superradiance produced by the corresponding $W$ state. While we derive our results for two-photon superradiance for arbitrary generalized $W$ states we focus in this paper on systems in doubly excited $W$ states; the outcomes for arbitrary $W$ states with more than two excitations  are presented in the Supplementary Information section.

Note that bunching in the radiation of generalized $W$ states can be explained semi-classically. However, the phenomenon of superbunching as well as the emission of nonclassical light, first demonstrated in 1977 \cite{Mandel(1977)}, can only be understood in a quantum mechanical description \cite{Mandel(1977),Paul(1982)}. The latter is a feature arising from the light's particle nature where photon fluctuations become smaller than for coherent light. Demonstrating nonclassicality in the light of arbitrary $W$ states thus directly leads to a manifestation of the particular quantum mechanical characteristics of these superradiant states. We finally discuss also the spatial cross correlations of photons in two-photon superradiance. Here, likewise, superbunching and nonclassicality can be observed.

We note that recent experiments by Jahnke et al. with quantum dots in a cavity have already reported the observation of superbunching \cite{Auffeves(2011),Jahnke(tbp)}. 
In this paper we bring out for a simple model system in free space the reasons for the appearance of this phenomenon,
a curio which does not commonly occur, the squeezed vacuum being one of the rare examples \cite{Boitier(2011)}. 

\section*{Results}

\label{sec:system}

\begin{figure}[t]
	\centering
		\includegraphics[width=0.55\textwidth]{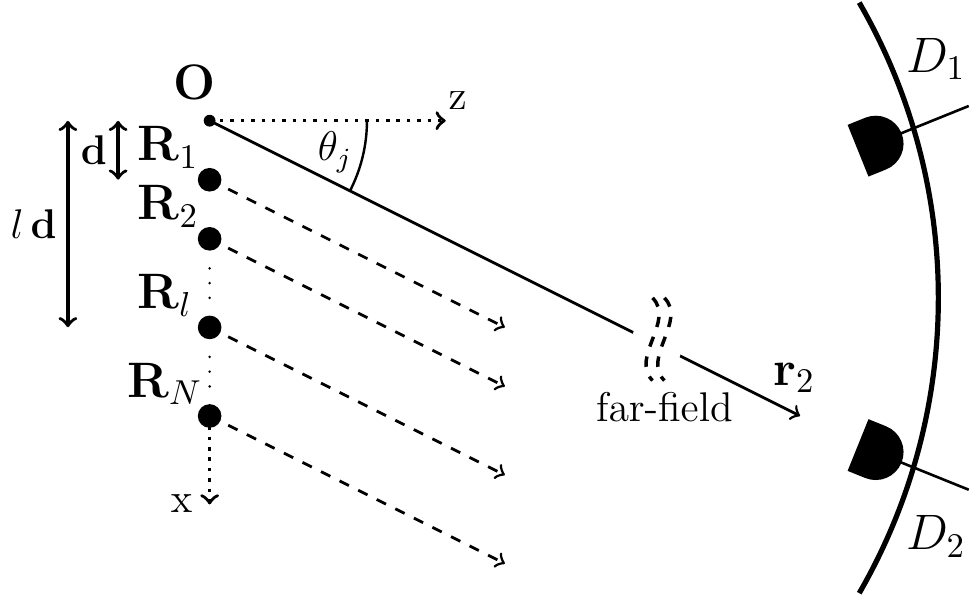}
\caption{Scheme of considered setup: $N$ two-level atoms are aligned on the x-axis, whereby neighboring atoms are separated by a distance $d$. The intensity in the far field $I(\vec{r}_{1})$ is measured by a single detector at position $\vec{r}_{1}$, whereas the second order correlation function $G^{(2)}(\vec{r}_{1},\vec{r}_{2})$ is measured by two detectors at $\vec{r}_{1}$ and $\vec{r}_{2}$.}
\label{fig:detection_scheme}
\end{figure}

To focus on the key aspects of two-photon superradiance we consider a linear system of $N$ equidistantly aligned identical emitters, e.g., atoms or ions 
with upper state $\ket{e_{l}}$ and ground state $\ket{g_{l}}$, $l =  1, \ldots , N$, trapped in a linear arrangement \cite{Meschede(2006),Wineland(2008),Blatt(2011),Rauschenbeutel(2014)} at positions ${\vec R}_{l}$
with spacing $d \gg \lambda$ such that the dipole-dipole coupling between the particles can be neglected (see Fig.~\ref{fig:detection_scheme}). The atoms are assumed to be prepared initially in a generalized $W$ state with $n_{e}$ excitations, i.e., in the state $\ket{W_{n_{e},N}}=\binom{N}{n_{e}}^{-\frac{1}{2}}\sum_{\{\alpha_{l}\}=\mathcal{P} {\{l\}}} \prod_{i=1}^{n_{e}}\ket{e_{\alpha_{i}}}\prod_{i=n_{e}+1}^{N}\ket{g_{\alpha_{i}}} $, 
where $\mathcal{P} {\{l\}}$ denotes all permutations of the set of atoms $\{l\}=\{1,2,\hdots,N\}$. In what follows we study the second order correlation functions at equal times emitted by the atoms in the described $W$ state.
To this end two detectors are placed at positions $\vec{r}_{1}$ and $\vec{r}_{2}$ in the far field  each measuring a single photon coincidentally, i.e., within a small time window much smaller than the lifetime of the upper state. 
To simplify the calculations we suppose that the emitters and the detectors are in one plane and that the atomic dipole moments of the transition $\ket{e_{l}} \rightarrow \ket{g_{l}}$ are oriented perpendicular to this plane (see Fig.~\ref{fig:detection_scheme}). 

Due to the far field condition and therefore the inability to identify the individual photon sources, the electric field operator at ${\vec r}_j$ takes the form \cite{Thiel(07)}
$\left[ \hat{E}^{(-)}(\vec{r}_j) \right]^\dagger \! = \hat{E}^{(+)}(\vec{r}_j) \sim \sum_{l=1}^{N} e^{-i\,\varphi_{lj}} \;\hat{s}^{-}_l = \sum_{l=1}^{N} e^{-i\,l\,\delta_{j}} \;\hat{s}^{-}_l$, 
where $\hat{s}_{l}^{-}=\ket{g_{l}}\!\bra{e_{l}}$  is the atomic lowering operator for atom $l$, and $\varphi_{lj}  = - k \, \frac{\vec{r}_j\cdot\vec{R}_l}{r_j} = l\, k\, d \, \sin\theta_{j} \cos\phi_j =l\,\delta_{j}$ the relative optical phase accumulated by a photon emitted by source $l$ 
and recorded by detector $j$ 
with respect to a photon emitted at the origin.
Note that the field operators have been chosen dimensionless as all dimension defining prefactors cancel out in the normalized correlation functions. 
The first and second order spatial correlation functions at equal times are defined as \cite{Glauber(1963)Quantum} $G^{(1)}(\vec{r}_{1})=\left<\hat{E}^{(-)}(\vec{r}_{1})\hat{E}^{(+)}(\vec{r}_{1})\right>$ and $G^{(2)}(\vec{r}_{1},\vec{r}_{2})=\left<\hat{E}^{(-)}(\vec{r}_{1})\hat{E}^{(-)}(\vec{r}_{2})\hat{E}^{(+)}(\vec{r}_{2})\hat{E}^{(+)}(\vec{r}_{1})\right> $, respectively, 
where $G^{(1)}(\vec{r}_{1})$ is proportional to the mean intensity of the emitted radiation, i.e.,  $G^{(1)}(\vec{r}_{1})\sim I(\vec{r}_{1}) $. To compare the photon statistics of various systems radiating with different intensities we further introduce the normalized second order correlation function \cite{Glauber(1963)Quantum} $g^{(2)}(\vec{r}_{1},\vec{r}_{2})=G^{(2)}(\vec{r}_{1},\vec{r}_{2})/(G^{(1)}(\vec{r}_{1})\,G^{(1)}(\vec{r}_{2}))$.

For the state $\ket{W_{n_{e},N}}$ the first order correlation function in the configuration of Fig.~\ref{fig:detection_scheme} has been calculated \cite{Wiegner(2011)PRA} to 
\begin{equation}
\begin{aligned}
	G^{(1)}_{n_{e},N}(\delta_{1})&= \frac{n_{e}(n_{e}-1)}{N-1} + \frac{N\, n_{e}(N-n_{e})}{N-1} \chi^{2}(\delta_{1}) \ ,
\label{eq:intdefinition}
\end{aligned}
\end{equation}
where $\chi(x)=\frac{\sin(\frac{Nx}{2})}{N\sin(\frac{x}{2})}$ corresponds to the normalized far-field intensity distribution of a coherently illuminated $N$-slit grating \cite{BORN(1999)}.
As it is well-known from classical optics, the distribution  $\chi^{2}(x)$ is strongly peaked in particular directions. The fact that $\chi^{2}(x)$ appears in the context of spontaneous emission of atoms in
generalized $W$ states as in Eq.~(\ref{eq:intdefinition}) has been coined by Dicke \textit{spontaneous emission of coherent radiation} or simply \textit{superradiance}  \cite{Dicke(1954)}. Note that even though $\chi^{2}(x)$ displays pronounced maxima in certain directions the distribution assumes also very small values and even vanishes in other directions what has been interpreted as \textsl{subradiance} of the states $\ket{W_{n_{e},N}}$ \cite{Wiegner(2011)PRA}. 

From Eq.~(\ref{eq:intdefinition}) we find that the intensity distribution of the state $\ket{W_{1,N}}$ with only one excitation ($n_{e}=1$) simplifies to $G^{(1)}_{1,N}(\delta_{1})=N\chi^{2}(\delta_{1})$, with $G^{(1)}_{1,N}(\delta_{1}) \rightarrow N \ \ \text{for} \ \delta_{1}=0$,
displaying a maximal visibility $\mathcal{V}=1$ and a peak value $N$ times the intensity of a single atom.
This kind of single photon superradiance has been extensively studied in the past  \cite{Scully(2006),Scully(2007),Scully(2009)Super,Scully(2009),Svidzinsky(2010),Wiegner(2011)PRA,Kaiser(2012),Kaiser(2013),Rohlsberger(2013)}. Its particular superradiant characteristics have been shown to result from quantum path interferences occurring due to the particular interatomic correlations of the collective state $\ket{W_{1,N}}$ \cite{Wiegner(2011)PRA}.

As discussed below the strong correlations of the states $\ket{W_{n_{e},N}}$ may lead to photon-photon correlations with $g_{n_{e},N}^{(2)}(\delta_{1},\delta_{1})> 2$ as well as $g_{n_{e},N}^{(2)}(\delta_{1},\delta_{1})<1$, corresponding to superbunched as well as nonclassical light, respectively.
Note that from the form of  normalized second order correlation function $g_{n_{e},N}^{(2)}(\delta_{1},\delta_{2})$ it is obvious that bunching necessitates small intensities,  i.e., $G^{(1)}(\delta_{1}) \ll G^{(2)}(\delta_{1},\delta_{1})$, whereas nonclassical light requires small values of the two-photon correlation function, i.e., $G^{(2)}(\delta_{1},\delta_{1})\ll G^{(1)}(\delta_{1})$.

In what follows we study two-photon superradiance for the simplest form of generalized $W$ states, i.e., $W$ states  with only two excitations, as the main features of two-photon superradiance can already be observed for this configuration; two-photon superradiance for arbitrary generalized $W$ states $\ket{W_{n_{e},N}}$ with $n_{e} > 2$ is discussed in the Supplementary Information section. 

Note that an atomic ensemble in the state $\ket{W_{n_{e}=2,N}}$ can be prepared for example by using photon pairs generated in a down conversion process. When sent on a $50:50$ beam splitter the photon pair would then produce two photons at either of the two output ports of the beam splitter, i.e., if one port delivers zero photons then the other one has two photons \cite{HongOuMandel(1987)}. Assuming perfect detection efficiency, a photon pair not registered at one output port of the beam splitter and not registered at the other one after having passed the atomic ensemble would then herald the absorption of two photons by the atomic system.

For $n_{e}=2$ the normalized second order correlation function for the considered configuration of Fig.~\ref{fig:detection_scheme} takes the form
\begin{equation}
\begin{aligned}
	& g_{2,N}^{(2)}(\delta_{1},\delta_{2})= \frac{N(N-1)}{2} \frac{\left( N \chi(\delta_{1})\chi(\delta_{2}) - \chi(\delta_{1}+\delta_{2}) \right)^{2}}{\left(1 + N(N-2)\chi^2(\delta_{1})\right) \left(1 + N(N-2)\chi^2(\delta_{2})\right)} \ .
\label{eq:Gcompletene2}
\end{aligned}
\end{equation}
This expression will be investigated in detail in the following subsections. 

\subsection*{Superbunching in Two-photon Superradiance}
\label{sec:ne2d1d1}

In this section we investigate whether bunching in two-photon superradiance, in particular the phenomenon of superbunching with $g_{n_{e},N}^{(2)}(\delta_{1},\delta_{2})> 2$, can be observed in the radiation produced by states of the form $\ket{W_{n_{e}=2,N}}$.
We start to explore the photon-photon correlations with the two detectors placed at equal positions, i.e., with the two spontaneously emitted photons recorded in the same mode; photon-photon cross correlations are studied thereafter. 


\begin{figure}[t]
	\centering
		\includegraphics[width=0.8\textwidth]{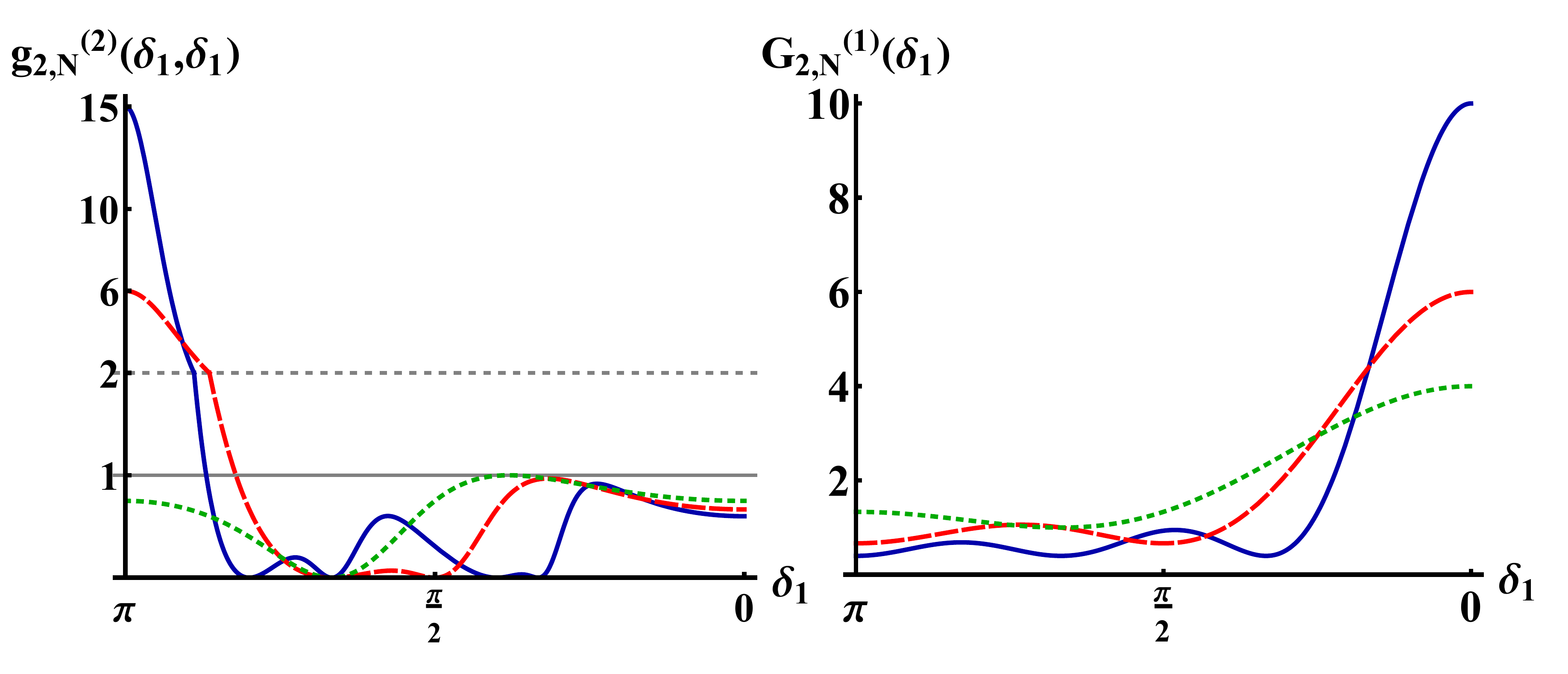}
	\caption{Second order correlation function $g^{(2)}_{2,N}(\delta_{1},\delta_{1})$ (left) and first order correlation function, i.e., the intensity distribution,  $G^{(1)}(\delta_{1})$ (right) of the radiation emitted by $N$ two-level atoms in the doubly excited $W$ state $\ket{W_{2,N}}$ for $N=3$ (dotted), $N=4$ (dashed), $N=6$ (solid). Superbunching is observed at $\delta_{1}=\pi$ for $N \geq 4$, while antibunching occurs at detector positions $\delta_{1}$ fulfilling $N \chi^{2}(\delta_{1})=\chi(2\delta_{1})$. For increasing $N$ superbunching becomes stronger, similar to single photon superradiance of the state $\ket{W_{1,N}}$. Comparing the two plots one can see that for high values of $G^{(1)}(\delta_{1})$ small values of $g^{(2)}_{2,N}(\delta_{1},\delta_{1})$ are obtained and vice versa.}
	\label{fig:ne2d1d1}
\end{figure}

According to Eq.~(\ref{eq:Gcompletene2}) the second order correlation function for the state $\ket{W_{2,N}}$ with two detectors placed at the same position takes the form
\begin{equation}
\begin{aligned}
	g^{(2)}_{2,N}(\delta_{1},\delta_{1})= \frac{ N(N-1)\, \left( N\,\chi^{2}(\delta_{1})-\chi(2\delta_{1}) \right)^{2}}{2\, \left( 1 + N(N-2)\, \chi^{2}(\delta_{1})  \right)^{2}} \ .
\label{eq:g2ne2d1d1}
\end{aligned}
\end{equation}

In order to access whether the system displays bunching for this configuration we have to search for values $g^{(2)}_{2,N}(\delta_{1},\delta_{1}) > 1$. Hitherto, 
we choose detector positions for which the values of the first order correlation function $G^{(1)}(\delta_{1})$ remain smaller than the unnormalized second order correlation function $G^{(2)}_{2,N}(\delta_{1},\delta_{1})$, i.e., locations where the intensity is low. This occurs for example at $\delta_{1}=\pi$, where $g^{(2)}_{2,N}(\delta_{1},\delta_{1})$ attains its maximal value (see Fig.~\ref{fig:ne2d1d1}). 
To investigate this outcome quantitatively we have to study the case of even and odd $N$ separately since $\chi(\pi)$ and $\chi(2\pi)$ yield different results in these two cases \cite{note(1)}.

For an even number $N$ of atoms one obtains the following two identities $\chi(\pi)=0$, $\chi(2\pi)=-1$,
from which we deduce $g^{(2)}_{2,N_{even}}(\pi,\pi)=N(N-1)/2$, 
leading to bunched light in case that $N_{even}\geq 4$ (see Fig.~\ref{fig:ne2d1d1}). In fact, for $N_{even}\geq 4$, we even obtain superbunching, i.e., $g^{(2)}_{2,N_{even}}(\pi,\pi) > 2$, what surpasses the maximum value achievable with classical light. 
Note that 
$g^{(2)}_{2,N_{even}}(\pi,\pi)$ as a function of $N$ has in principal no upper limit as it increases $\sim N^{2}$ for $N\gg1$. This means that we can produce principally unlimited values of superbunching if we add more and more atoms in the ground state to the system \cite{Auffeves(2011),Jahnke(tbp)}.

In case of odd $N$ the above identities read $\chi(\pi)=\pm 1/N$, $\chi(2\pi)= +1$.
what leads to the a maximal value of the second oder correlation function of $g^{(2)}_{2,N_{odd}}(\pi,\pi)=N(N-1)/8$.
Here we obtain bunched and superbunched light in case that $N_{odd} \geq 5$. 
Again, as in the case of even $N$, there is no upper limit for the maximum value of the two-photon correlation function,
as once more we have $g^{(2)}_{2,N_{odd}}(\pi,\pi) \sim N^{2}$.

\subsection*{Nonclassicality in Two-photon Superradiance}
\label{sec:ne2d1d1anti}

In this section we investigate whether for the initial state $\ket{W_{2,N}}$ and two detectors at equal positions we can obtain nonclassical light, a sub-Poissonian photon statistics and antibunching in two-photon superradiance. 
As known from the radiation of a single atom \cite{Mandel(1977),Paul(1982)} a sub-Poissonian photon statistics derives from the discrete nature of the scattered radiation and can be explained only in a quantum mechanical treatment of resonance fluorescence. A complete vanishing of the second order correlation function at equal times for the state $\ket{W_{2,N}}$ indicates that $g^{(2)}_{2,N}(\delta_{1},\delta_{1}) = 0$ at $\tau = 0$ what proves true antibunching, whereas values $g^{(2)}_{2,N}(\delta_{1},\delta_{1}) < 1$ result from the nonclassical nature of the radiation scattered by the state $\ket{W_{2,N}}$.

Noting that the second order correlation function $g^{(2)}_{2,N}(\delta_{1},\delta_{1})$ displays a visibility of $\mathcal{V}=1$ (see Fig.~\ref{fig:ne2d1d1}) there must be indeed detector positions where $g^{(2)}_{2,N}(\delta_{1},\delta_{1}) = 0$.
More specifically, according to Eq.~($\ref{eq:g2ne2d1d1}$), the photon-photon correlation function vanishes independently of the atom number $N$ in case that the detectors are located at positions $\tilde{\delta}_{1}=a\frac{2\pi}{N}$, with $ a = 1,2, \ldots < N/2$, as in this case  $\chi(\tilde{\delta}_{1})=\chi(2 \tilde{\delta}_{1})=0$. Towards these positions the atomic system thus radiates photons which display complete antibunching \cite{note(2)}.
If the two detectors are located at $\delta_{1}=0$ we have $g^{(2)}_{2,N}(\delta_{1},\delta_{1}) < 1$ as long as $N\geq 3$, as in this case the photon-photon correlation function takes the form $g^{(2)}_{2,N}(0,0)=N/(2(N-1))$ (cf. Eq.~(\ref{eq:g2ne2d1d1})).
Hence, in those directions the atomic system emits  nonclassical light with photon number fluctuations smaller than those for coherent light.
\begin{figure}[t]
	\centering
		\includegraphics[width=0.8\textwidth]{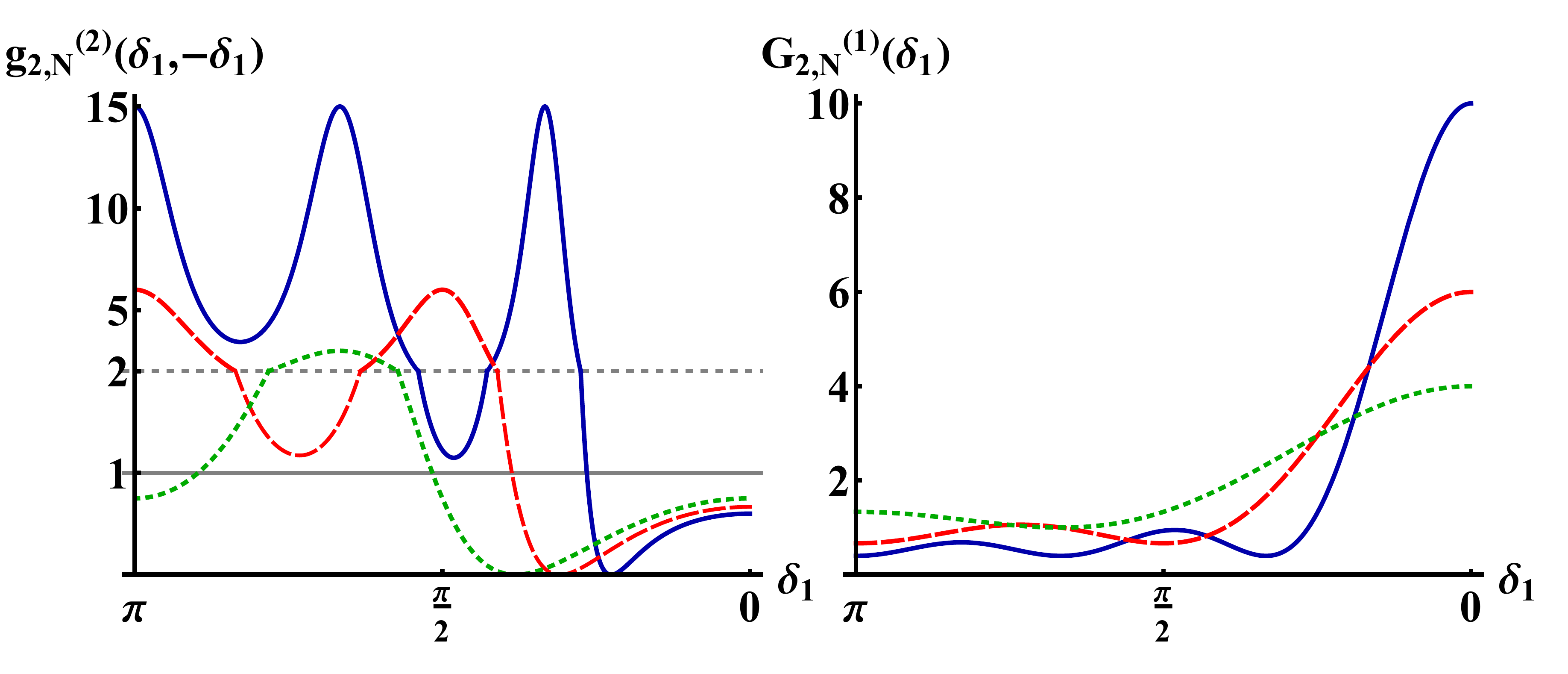}
	\caption{Left: Second order correlation function $g_{n_{e},N}^{(2)}(\delta_{1},\delta_{2})$ for $n_{e}=2$, $\delta_{2}=-\delta_{1}$ and $N=3$ (dotted), $N=4$ (dashed), $N=6$ (solid). The function displays both superbunching and antibunching with a maximal visibilities of $\mathcal{V}=1$. It can be seen that the superbunching effect increases with increasing $N$, while for $\delta_{1}=0$ the second order correlation function converges to $1/2$. Right: First order correlation function $G^{(1)}_{2,N}(\delta_{1})$ for $N=3$ (dotted), $N=4$ (dashed), $N=6$ (solid). From the figure it can be seen that for small values of $G^{(1)}_{2,N}(\delta_{1})$ high values of $g_{n_{e},N}^{(2)}(\delta_{1},\delta_{2})$ are obtained. In contrast to the case $\delta_{1}=\delta_{2}$ several superbunching peaks occur.}
	\label{fig:ne2d1minusd1}
\end{figure}

\subsection*{Two-photon Cross Correlations}
\label{sec:crosscorrelations}

Finally we study  the spatial cross correlations in two-photon superradiance for atoms in the state $\ket{W_{2,N}}$, i.e, the behavior of the second order correlation function $g^{(2)}_{2,N}(\delta_{1}, \delta_{2})$ in case that the scattered photons are recorded at different positions $\delta_{1} \neq \delta_{2}$. We start to explore the particular configuration of counter-propagating detectors, i.e., detectors at positions $\delta_{1}= - \delta_{2}$, followed by the case where $\delta_{2}$ is fixed and only $\delta_{1}$ is varied. 

For $\delta_{1}= -\delta_{2}$ (i.e., where we choose $\theta_1 = \theta_2$, $\phi_2 = 0$ and $\phi_1 = \pi$),
the second order correlation function reads (cf. Eq.~(\ref{eq:Gcompletene2}))
\begin{equation}
		g^{(2)}_{2,N}(\delta_{1},-\delta_{1})= \frac{N (N-1)\, \left(N\, \chi^{2}(\delta_{1}) - 1 \right)^{2} }{ 2\, \left( 1 + N(N-2)\, \chi^{2}(\delta_{1}) \right)^{2} } \ .
\label{eq:g2ne2d1minusd1}
\end{equation}
To determine the possibilities for superbunching in cross correlations of the scattered photons we look for the maximum of Eq.~(\ref{eq:g2ne2d1minusd1}). This is attained at $\bar{\delta}_{1}=a\frac{2\pi}{N}$, with $a = 1, \ldots, N/2$, in which case $\chi(\delta_{1})$ vanishes and  the second order correlation function reads $g^{(2)}_{2,N}(\bar{\delta},-\bar{\delta})=N(N-1)/2$.
This result is principally identical to $g^{(2)}_{2,N_{even}}(\pi,\pi)$. However, in the case of counter-propagating detectors it is valid for arbitrary $N$, i.e., for $N$ even or odd. Here, the threshold for superbunching is exceeded if $N\geq 3$ (see Fig.~\ref{fig:ne2d1minusd1}).
In the same configuration also 
complete antibunching 
can be obtained. This occurs for $\hat{\delta}_{1}$ such that $\chi^{2}(\hat{\delta}_{1}) = 1/N$ (cf. Eq.~(\ref{eq:g2ne2d1minusd1})), in which case the photon correlation function vanishes identically, i.e., $g^{(2)}_{2,N}(\hat{\delta}_{1},-\hat{\delta}_{1})= 0$. 

\begin{figure}[t]
	\centering
		\includegraphics[width=0.8\textwidth]{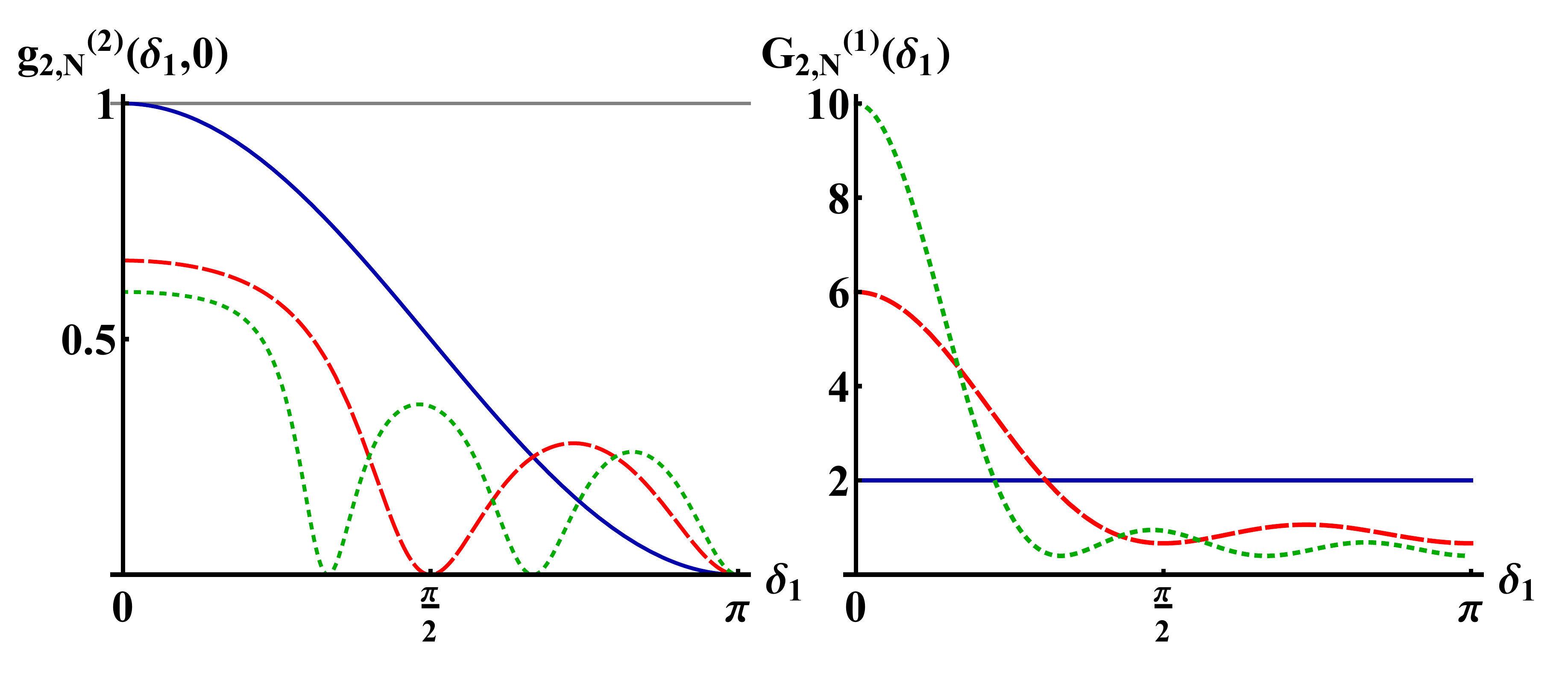}
	\caption{Left: Second order correlation function $g_{n_{e},N}^{(2)}(\delta_{1},\delta_{2})$ for $n_{e}=2$ where one detector is fixed at $\delta_{2}=0$. For all $N$ (in the plot: $N=2$ (solid), $N=4$ (dashed), $N=6$ (dotted)) the two-photon correlation function remains smaller than one for any $\delta_{1}\neq 0$. For $\delta_{1}=0$ the maximal value of one is attained only in the case $N=2$. Right: First order correlation function $G^{(1)}_{2,N}(\delta_{1})$ for $N=2$ (solid), $N=4$ (dashed), $N=6$ (dotted).}
	\label{fig:ne2d10}
\end{figure}

Another interesting configuration is the case when $\delta_{2}$ is fixed and only $\delta_{1}$ is varied. If we fix $\delta_{2}=0$ the photon-photon cross correlation function takes the form 
(cf. Eq.~(\ref{eq:Gcompletene2}))
\begin{equation}
	g^{(2)}_{2,N}(\delta_{1},0) = \frac{N(N-1)\chi^{2}(\delta_{1})}{2 + 2N(N-2)\chi^{2}(\delta_{1})} \ ,
	\label{eq:g2ne2delta10}
\end{equation}
which can maximally take a value of $1$ (for $N=2$ and $\delta_{1} = 0$), whereas for  $N > 2$ the cross correlation function remains always $< 1$. In Fig.~\ref{fig:ne2d10} the corresponding photon-photon cross correlations are shown for $N = 2, 4, 6$ for the entire range $\delta_{1} \in [0, \pi]$. It can be seen that by increasing $N$ the second order correlation function $g^{(2)}_{2,N}(\delta_{1},0)$ decreases in absolute values. The reason is the following: since $G^{(2)}_{2,N}(\delta_{1},0)$ and $G^{(1)}_{2,N}(\delta_{1})$ depend on $\chi(\delta_{1})$ in a similar way the overall behavior of $g^{(2)}_{2,N}(\delta_{1},0)$ is determined by the prefactors of $\chi(\delta_{1})$ in the numerator and denominator of $g^{(2)}_{2,N}(\delta_{1},0)$, the ratio of which decreases with increasing $N$ and converges  to $1/2$ for $N \gg 1$.

\section*{Discussion}

In conclusion we investigated for a prototype ensemble of $N$ identical non-interacting 
two-level atoms prepared in collective
superradiant generalized $W$-states with $n_{e}$ excitations the particular quantum statistical properties of the emitted radiation. Such investigations require the collective system to have at least two excitations as we explore the photon-photon correlations of the scattered light. We derived conditions for which the atomic system emits bunched and even superbunched light, as well as nonclassial and antibunched radiation. Here, superbunching refers to values of the normalized second order correlation function $g^{(2)}_{2,N}(\delta_{1}, \delta_{1}) > 2$ and antibunching to values $g^{(2)}_{2,N}(\delta_{1}, \delta_{1}) = 0$. In some cases the results were obtained under the condition that the number of atoms in the ensemble exceed a certain threshold. For example, the smallest number of atoms producing superbunching in the state $\ket{W_{2,N}}$ is $N=3$;
similar results were derived for $N$ atoms with arbitrary number of excitations, 
as shown in the Supplementary Information section. 
Note that in coherently driven atomic systems superbunching can be observed already for $N = 2$. For example, in \cite{Skornia(2001)} it was demonstrated that arbitrarily high values of $g^{(2)}_{N = 2}(\delta_{1}, \delta_{1})$ can be produced for $\delta_{1} = \pi$ in case that the two atoms are very weakly excited, as in this case  $g^{(2)}_{N = 2}(\pi, \pi) \sim 1/\Omega^4$, where $\Omega$ is the Rabi frequency (in units of the spontaneous decay rate $\gamma$). The effect is even stronger if the two atoms are subject to a strong dipole-dipole interaction as in dipole blockade systems; here $g^{(2)}_{N = 2}(\pi, \pi) \sim \delta_0^2/\Omega^4$ where $\delta_0$ is the level shift of the doubly excited state (again in units of $\gamma$) \cite{Bastin(2010)}.
In the last part of the paper we finally investigated the spatial cross correlations in two-photon superradiance, i.e., the second order correlation functions $g^{(2)}_{2,N}(\delta_{1}, \delta_{2})$ for detector positions $\delta_{1} \neq \delta_{2}$. Here, again, positions were found where $g^{(2)}_{2,N}(\delta_{1}, \delta_{2}) > 2$ and $g^{(2)}_{2,N}(\delta_{1}, \delta_{2}) = 0$, corresponding in this case to superbunching and antibunching of the cross correlations of the scattered photons.




\section*{Acknowledgements}

GSA and JvZ thank Frank Jahnke for fruitful discussions and the disclosure of documents prior to publication. DB and JvZ warmly thank the hospitality of GSA at various stays at the Oklahoma State University. DB thanks the Friedrich-Alexander Universit\"at Erlangen-N\"urnberg for a travel grant ``Hochschule International'' and the German Academic Exchange Service for a PROMOS travel grant. The authors gratefully acknowledge funding by the Erlangen Graduate School in Advanced Optical Technologies (SAOT) by the German Research Foundation (DFG) in the framework of the German excellence initiative.

\section*{Author contributions statement}

DB did all the calculations while the ideas were formulated by GSA. JvZ helped with thoughts in possible experimental implementation. JvZ primarily wrote the manuscript for publication which was reviewed by both DB and GSA.

\section*{Additional information}

The authors declare no competing financial interests.



\section*{Supplementary material}

# Supplementary Information
# Superbunching and Nonclassicality as new Hallmarks of Superradiance

Daniel Bhatti[1,2,*], Joachim von Zanthier[3], and Girish S. Agarwal[2]

[1]Institut für Optik, Information und Photonik, Universität Erlangen-Nürnberg, 91058 Erlangen, Germany
[2]Department of Physics, Oklahoma State University, Stillwater, OK, USA
[3]Erlangen Graduate School in Advanced Optical Technologies (SAOT), Universität Erlangen-Nürnberg, 91052 Erlangen, Germany
[*]daniel.bhatti@fau.de

## Photon-photon correlations for atoms in arbitrary generalized $W$ states

In this Supplementary Information section we calculate the second order correlation function for $N$ atoms in a generalized $W$ state $|W_{n_e,N}\rangle$, i.e., with an arbitrary number of excitations $n_e$. For this purpose the correlation function of second order can be written as

$$G^{(2)}_{n_e,N}(\delta_1,\delta_2) = \binom{N}{n_e}^{-1} \sum_{|\Psi_f\rangle} |\sum_{l_1,l_2 \in \{\Psi_f\}} e^{i(l_1\delta_1 + l_2\delta_2)}|^2 \,, \quad (S1)$$

with $|\Psi_f\rangle$ denoting all possible final states. Each final state is characterized by a particular set of $N - n_e + 2$ integers $\{\Psi_f\} \subseteq \{1,\ldots,N\}$ representing all sources, that could have emitted the detected two photons. Since each source can emit only one photon, we have $l_1 \neq l_2$. Note that for $n_e = 2$ there is only one final state and the sum over $|\Psi_f\rangle$ vanishes. From Eq. (S1) one can find a combinatorial solution to the problem

$$G^{(2)}_{n_e,N}(\delta_1,\delta_2) = \binom{N}{n_e}^{-1} \sum_{\substack{l'_1,l'_2,l_1,l_2=1 \\ l'_1 \neq l'_2, l_1 \neq l_2}}^{N-1} A_{l'_1,l'_2,l_1,l_2} e^{-i(l'_1\delta_1 + l'_2\delta_2)} e^{i(l_1\delta_1 + l_2\delta_2)} \,, \quad (S2)$$

where each term requires a statistical loading $A_{l'_1,l'_2,l_1,l_2}$ depending on the phase factors $l'_1 \neq l'_2$ and $l_1 \neq l_2$. With $m = (0 \times 2), (1 \times 2), (2 \times 2)$ of these factors being equal the statistical loading corresponds to

$$A_{l'_1,l'_2,l_1,l_2} = \binom{N-4+\frac{m}{2}}{n_e - 2} \,, \quad (S3)$$

since there are $n_e - 2$ excitations that can be distributed over $N - 4 + m/2$ atoms. This leads to (cf. Eq. (S2))

$$G^{(2)}_{n_e,N}(\delta_1,\delta_2) = \binom{N}{n_e}^{-1} \left[ \binom{N-4}{n_e-2} B_{m=0} + \binom{N-3}{n_e-2} B_{m=2} + \binom{N-2}{n_e-2} B_{m=4} \right] \,. \quad (S4)$$

We now calculate the different terms $B_m$ of $G^{(2)}_{n_e,N}(\delta_1,\delta_2)$ separately, starting with the case $m=4$

$$\begin{aligned} B_{m=4} &= \sum_{L \neq K} e^{-i(L\delta_1 + K\delta_2)} e^{i(L\delta_1 + K\delta_2)} + \sum_{L \neq K} e^{-i(L\delta_1 + K\delta_2)} e^{i(K\delta_1 + L\delta_2)} \\ &= \sum_{L,K=1}^{N} e^{-i(L\delta_1 + K\delta_2)} e^{i(L\delta_1 + K\delta_2)} - \sum_{L=1}^{N} e^{-iL(\delta_1+\delta_2)} e^{iL(\delta_1+\delta_2)} \\ &+ \sum_{L,K=1}^{N} e^{-i(L\delta_1 + K\delta_2)} e^{i(K\delta_1 + L\delta_2)} - \sum_{L=1}^{N} e^{-iL(\delta_1+\delta_2)} e^{iL(\delta_1+\delta_2)} \\ &= N^2 - 2N + N^2 \chi^2(\delta_1 - \delta_2) \,, \end{aligned} \quad (S5)$$

where in the last step the following identity has been used

$$\sum_{K,L=1}^{N} e^{iK\delta} e^{-iL\delta} = \frac{\sin^2(\frac{N\delta}{2})}{\sin^2(\frac{\delta}{2})} = N^2 \chi^2(\delta) \ . \tag{S6}$$

$B_{m=4}$ (cf. Eq. (S5)) can now be used to calculate $B_{m=2}$

$$\begin{aligned}
B_{m=2} &= \sum_{L \neq l'_2 \neq l_2 \neq L} e^{-i(L\delta_1 + l'_2\delta_2)} e^{i(L\delta_1 + l_2\delta_2)} + \sum_{L \neq l'_1 \neq l_1 \neq L} e^{-i(l'_1\delta_1 + L\delta_2)} e^{i(l_1\delta_1 + L\delta_2)} \\
&\quad + \sum_{L \neq l'_2 \neq l_1 \neq L} e^{-i(L\delta_1 + l'_2\delta_2)} e^{i(l_1\delta_1 + L\delta_2)} + \sum_{L \neq l'_1 \neq l_2 \neq L} e^{-i(l'_1\delta_1 + L\delta_2)} e^{i(L\delta_1 + l_2\delta_2)} \\
&= \sum_{L \neq l'_2, l_2} e^{-i(L\delta_1 + l'_2\delta_2)} e^{i(L\delta_1 + l_2\delta_2)} + \sum_{L \neq l'_1, l_1} e^{-i(l'_1\delta_1 + L\delta_2)} e^{i(l_1\delta_1 + L\delta_2)} \\
&\quad + \sum_{L \neq l'_2, l_1} e^{-i(L\delta_1 + l'_2\delta_2)} e^{i(l_1\delta_1 + L\delta_2)} + \sum_{L \neq l'_1, l_2} e^{-i(l'_1\delta_1 + L\delta_2)} e^{i(L\delta_1 + l_2\delta_2)} - 2B_{m=4} \ .
\end{aligned} \tag{S7}$$

The two terms

$$\begin{aligned}
\sum_{L \neq l'_2, l_2} e^{-i(L\delta_1 + l'_2\delta_2)} e^{i(L\delta_1 + l_2\delta_2)} &= \sum_{L, l'_2, l_2=1}^{N} e^{-i(L\delta_1 + l'_2\delta_2)} e^{i(L\delta_1 + l_2\delta_2)} - \sum_{L, l'_2=1}^{N} e^{-i(L\delta_1 + l'_2\delta_2)} e^{iL(\delta_1 + \delta_2)} \\
&\quad - \sum_{L, l_2=1}^{N} e^{-iL(\delta_1 + \delta_2)} e^{i(L\delta_1 + l_2\delta_2)} + \sum_{L=1}^{N} e^{-iL(\delta_1 + \delta_2)} e^{iL(\delta_1 + \delta_2)} \\
&= N + N^2(N-2)\chi^2(\delta_2) \ ,
\end{aligned} \tag{S8}$$

and

$$\begin{aligned}
\sum_{L \neq l'_2, l_2} e^{-i(L\delta_1 + l'_2\delta_2)} e^{i(l_1\delta_1 + L\delta_2)} &= \sum_{L, l'_2, l_2=1}^{N} e^{-i(L\delta_1 + l'_2\delta_2)} e^{i(l_1\delta_1 + L\delta_2)} - \sum_{L, l_1=1}^{N} e^{-iL(\delta_1 + \delta_2)} e^{i(l_1\delta_1 + L\delta_2)} \\
&\quad - \sum_{L, l'_2=1}^{N} e^{-i(L\delta_1 + l'_2\delta_2)} e^{iL(\delta_1 + \delta_2)} + \sum_{L=1}^{N} e^{-iL(\delta_1 + \delta_2)} e^{iL(\delta_1 + \delta_2)} \\
&= N - N^2\chi^2(\delta_2) - N^2\chi^2(\delta_1) + \sum_{L, l_1, l_2=1}^{N} e^{iL(\delta_2 - \delta_1)} e^{il_1\delta_1} e^{-il_2\delta_2} \ ,
\end{aligned} \tag{S9}$$

lead to the solution of Eq. (S7), which reads

$$B_{m=2} = -2N^2 + 8N + N^2(N-4)\chi^2(\delta_1) + N^2(N-4)\chi^2(\delta_2) - 2N^2\chi^2(\delta_1 - \delta_2) + 2N^3\chi(\delta_1)\chi(\delta_2)\chi(\delta_1 - \delta_2) \ , \tag{S10}$$

where the identity

$$\begin{aligned}
\sum_{L,l_1,l_2=1}^{N} e^{iL(\delta_2-\delta_1)} e^{il_1\delta_1} e^{-il_2\delta_2} + \sum_{L,l_1,l_2=1}^{N} e^{iL(\delta_1-\delta_2)} e^{il_1\delta_2} e^{-il_2\delta_1} &= 2\sum_{L,l_1,l_2=1}^{N} \cos(L(\delta_1 - \delta_2) + l_1\delta_1 - l_2\delta_2) \\
&= 2N^3\chi(\delta_1)\chi(\delta_2)\chi(\delta_1 - \delta_2) \ ,
\end{aligned} \tag{S11}$$

has been used.

The last term $B_{m=0}$ of Eq. (S4) can now be calculated

$$\begin{aligned}
B_{m=0} &= \sum_{l'_1 \neq l'_2; \ l_1 \neq l_2} e^{-i(l'_1\delta_1 + l'_2\delta_2)} e^{i(l_1\delta_1 + l_2\delta_2)} - B_{m=2} - B_{m=4} \\
&= \sum_{l'_1, l'_2, l_1, l_2=1}^{N} e^{-i(l'_1\delta_1 + l'_2\delta_2)} e^{i(l_1\delta_1 + l_2\delta_2)} - \sum_{l'_1, l'_2, L=1}^{N} e^{-i(l'_1\delta_1 + l'_2\delta_2)} e^{iL(\delta_1 + \delta_2)} \\
&\quad - \sum_{L, l_1, l_2=1}^{N} e^{-iL(\delta_1 + \delta_2)} e^{i(l_1\delta_1 + l_2\delta_2)} + \sum_{L, K=1}^{N} e^{-iL(\delta_1 + \delta_2)} e^{iK(\delta_1 + \delta_2)} - B_{m=2} - B_{m=4} \\
&= N^2 - 6N - N^2(N-4)\chi^2(\delta_1) - N^2(N-4)\chi^2(\delta_2) + N^4\chi^2(\delta_1)\chi^2(\delta_2) + N^2\chi^2(\delta_1 + \delta_2) + N^2\chi^2(\delta_1 - \delta_2) \\
&\quad - 2N^3\chi(\delta_1)\chi(\delta_2)\chi(\delta_1 + \delta_2) - 2N^3\chi(\delta_1)\chi(\delta_2)\chi(\delta_1 - \delta_2) \ ,
\end{aligned} \tag{S12}$$

where identities (S6) and (S11) have been used and $B_{m=2}$ (cf. Eq. (S10)) and $B_{m=4}$ (cf. Eq. (S5)) have been inserted.

Summing over all different terms the correlation function finally takes the form (cf. Eq. (S4))

$$
\begin{aligned}
G^{(2)}_{n_e,N}(\delta_1,\delta_2) = \binom{N}{n_e}^{-1} \Bigg[ & \binom{N-4}{n_e-2}\Big[ N^2 - 6N - N^2(N-4)\chi^2(\delta_1) - N^2(N-4)\chi^2(\delta_2) \\
& + N^4\chi^2(\delta_1)\chi^2(\delta_2) + N^2\chi^2(\delta_1+\delta_2) + N^2\chi^2(\delta_1-\delta_2) \\
& - 2N^3\chi(\delta_1)\chi(\delta_2)\chi(\delta_1+\delta_2) - 2N^3\chi(\delta_1)\chi(\delta_2)\chi(\delta_1-\delta_2) \Big] \\
& + \binom{N-3}{n_e-2}\Big[ -2N^2 + 8N + N^2(N-4)\chi^2(\delta_1) + N^2(N-4)\chi^2(\delta_2) \\
& - 2N^2\chi^2(\delta_1-\delta_2) + 2N^3\chi(\delta_1)\chi(\delta_2)\chi(\delta_1-\delta_2) \Big] \\
& + \binom{N-2}{n_e-2}\Big[ N^2 - 2N + N^2\chi^2(\delta_1-\delta_2) \Big] \Bigg].
\end{aligned}
$$
(S13)

In the following two sections we investigate the second order correlation functions obtained in case of using generalized $W$-states $|W_{n_e,N}\rangle$ with higher numbers of excitations $n_e > 2$.

## Superbunching in the radiation of $N$ atoms in arbitrary generalized $W$ states

Similar to the case of $n_e = 2$ we want to investigate whether bunching and superbunching can be observed in two-photon superradiance from generalized $W$ states of higher excitations $|W_{n_e>2,N}\rangle$. We start to explore the photon-photon correlations with the two detectors placed at equal positions. In this case the normalized correlation function reads (cf. Eq. (S13))

$$
\begin{aligned}
g^{(2)}_{n_e,N}(\delta_1,\delta_1) = \mathcal{N}\binom{N}{n_e}^{-1}\Bigg[& \binom{N-2}{n_e-2}2N[N-1] - \binom{N-3}{n_e-2}4N(N-2)\left[1-N\chi^2(\delta_1)\right] \\
& + \binom{N-4}{n_e-2}\left[2N(N-3) - 4N^2(N-2)\chi^2(\delta_1) + \left(N\chi(2\delta_1) - N^2\chi^2(\delta_1)\right)^2\right]\Bigg],
\end{aligned}
$$
(S14)

with the normalization $\mathcal{N} = (G^{(1)}_{n_e,N}(\delta_1))^{-2}$ (cf. Eq (1)).

To access whether the correlation function displays bunching we investigate the conditions for $g^{(2)}_{n_e,N} > 1$. As in the case $n_e = 2$, we choose the position $\delta_1 = \pi$ where $g^{(2)}_{n_e,N}$ attains its maximal value and the normalizing intensity is small (see Fig. S1). Since $\chi(\pi)$ and $\chi(2\pi)$ yield different results for even $N$ and odd $N$ these two cases have to be investigated separately.

In case of an even $N$ the second-order correlation function reads (cf. Eq. (S14))

$$
g^{(2)}_{n_e,N_{even}}(\pi,\pi) = \frac{(N-1)(6+N^2+N-9n_e-2Nn_e+3n_e^2)}{(N-3)n_e(n_e-1)}.
$$
(S15)

For large $N \gg n_e$ we can, as in case of $n_e = 2$, produce principally unlimited values of superbunching since there is no upper limit to $g^{(2)}_{n_e,N_{even}}(\pi,\pi)$ when adding more and more atoms in the ground state.[1,2] In this limit the correlation function simplifies to

$$
g^{(2)}_{n_e,N_{even}}(\pi,\pi) \sim \frac{N^2}{n_e(n_e-1)},
$$
(S16)

clearly exceeding both the threshold for bunching ($g^{(2)}_{n_e,N} = 1$) as well as the threshold for superbunching ($g^{(2)}_{n_e,N} = 2$), whereby the fastest growth is obtained for $n_e = 2$. For example, for $n_e = 3$, Eq. (S15) is given by

$$
g^{(2)}_{3,N_{even}}(\pi,\pi) = \frac{(N-1)(N-2)}{6},
$$
(S17)

leading to bunched and superbunched light for $N_{even} \geq 6$.

In case of odd $N$, the following maximal value of the second order correlation function can be calculated (cf. Eq. (S14))

$$
g^{(2)}_{n_e,N_{odd}}(\pi,\pi) = \frac{N(n_e-1)(N^2+N-2Nn_e-5n_e+3n_e^2)}{n_e^3(N-2)}.
$$
(S18)

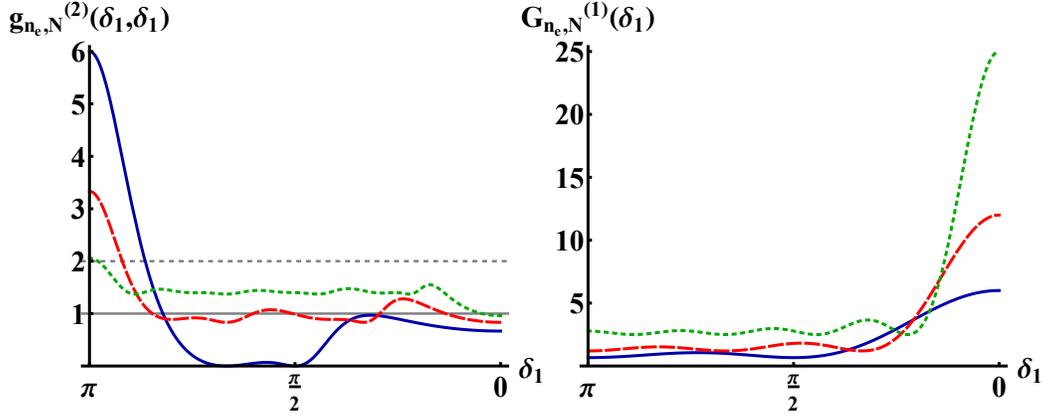

**Figure S1.** Second order correlation function $g^{(2)}_{n_e,N}(\delta_1,\delta_1)$ (left) and first order correlation function $G^{(1)}_{n_e,N}(\delta_1)$ (right) for $(n_e, N) = (2,4)$ (solid), $(3,6)$ (dashed), $(5,9)$ (dotted). For $\delta_1 = 0$ superbunching is observed for all displayed cases, whereas antibunching is obtained only for $n_e = 2$, independently of $N$ (see Eq. (S14)). For $\delta_1 = 0$ the second order correlation function displays nonclassicality in case that $2(n_e - 1) < N$ (see Eq. (S21)). Comparing the two plots one can see that for high values of $G^{(1)}_{n_e,N}(\delta_1)$ small values of $g^{(2)}_{n_e,N}(\delta_1,\delta_1)$ are obtained and vice versa.

From Eq. (S18) the behavior of the second order correlation function for $N \gg n_e$ can be derived

$$g^{(2)}_{n_e,N_{odd}}(\pi,\pi) \sim \frac{N^2(n_e-1)}{n_e^3} \,, \tag{S19}$$

which again may display bunching as well as superbunching, as there is no upper limit to Eq. (S19) when increasing $N$. For example, for $n_e = 3$ and odd $N$ the second order correlation function reads

$$g^{(2)}_{3,N_{odd}}(\pi,\pi) = \frac{2N(N^2 - 5N + 12)}{27(N-2)} \,, \tag{S20}$$

which shows bunching for $N_{odd} \geq 3$ and superbunching for $N_{odd} \geq 7$.

## Nonclassicality and antibunching in the radiation of $N$ atoms in arbitrary generalized $W$ states

We next investigate whether for initial arbitrary generalized $W$ states we can observe also nonclassical light and antibunching in two-photon superradiance. Again, we start to explore the photon-photon correlations with the two detectors placed at equal positions.

Note that a visibility of $\mathcal{V} = 1$ of the second order correlation function is obtained only for initially doubly excited states, what rules out true antibunching for $n_e \geq 3$ (see Eq. (S14)). However, for increasing $n_e$, we can always find $g^{(2)}_{n_e,N}(\delta_1,\delta_1) < 1$ at $\delta_1 = 0$, what can be explained physically since here the intensity attains its maximum $G^{(1)}_{n_e,N}(0) = n_e(N - n_e + 1)$.[3] To prove this we calculate from Eq. (S14)

$$g^{(2)}_{n_e,N}(0,0) = \frac{(n_e-1)(N-n_e+2)}{n_e(N-n_e+1)} \,, \tag{S21}$$

what shows that the atomic system emits nonclassical light with photon number fluctuations smaller than those for coherent light under the condition that $2(n_e - 1) < N$.

## Cross correlations in the radiation of $N$ atoms in arbitrary generalized $W$ states

Finally, we study the spatial cross correlations in two-photon superradiance for atoms in the generalized state $|W_{n_e,N}\rangle$, i.e, the behavior of the second order correlation function $g^{(2)}_{n_e,N}(\delta_1,\delta_2)$ in case that the scattered photons are recorded at different

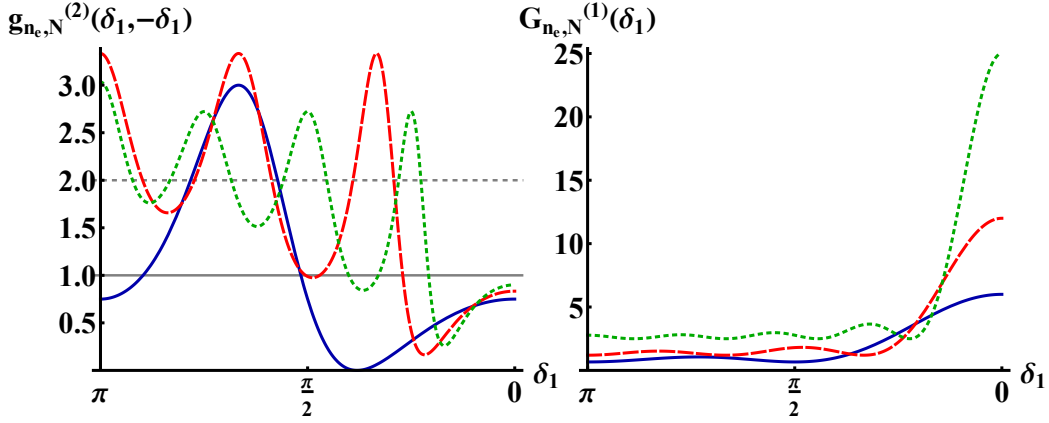

**Figure S2.** Second order correlation function $g^{(2)}_{n_e,N}(\delta_1,-\delta_1)$ (left) and first order correlation function $G^{(1)}_{n_e,N}(\delta_1)$ (right) for $(n_e,N) = (2,3)$ (solid), $(3,6)$ (dashed), $(4,8)$ (dotted). For a given $n_e$, the number of atoms $N$ has been chosen in such a way that superbunching is fulfilled for all positions $\tilde{\delta}_1 = a\frac{2\pi}{N}$, with $a = 1,2,\ldots < N/2$ (see Eq. (S24)). Again, antibunching is obtained only for $n_e = 2$, independently of $N$ (see Eq. (S22)). For $\delta_1 = 0$ the second order correlation function displays nonclassicality in case that $2(n_e - 1) < N$ (see Eq. (S21)).

positions $\delta_1 \neq \delta_2$. We start to explore the particular configuration of counter-propagating detectors, i.e., detectors at positions $\delta_1 = -\delta_2$, followed by the case where $\delta_2$ is fixed and only $\delta_1$ is varied.

For the case $\delta_1 = -\delta_2$, the second order correlation function simplifies to (cf. Eq. (S13))

$$g^{(2)}_{n_e,N}(\delta_1,-\delta_1) = \mathcal{N}\binom{N}{n_e}^{-1}\left[\binom{N-2}{n_e-2}N\left[N-2+N\chi^2(2\delta_1)\right]\right.$$
$$-\binom{N-3}{n_e-2}2N\left[(N-4)\left(1-N\chi^2(\delta_1)\right) + N\chi^2(2\delta_1) - N^2\chi^2(\delta_1)\chi(2\delta_1)\right] \quad (S22)$$
$$\left.+\binom{N-4}{n_e-2}\left[2N(N-3) - 4N^2(N-2)\chi^2(\delta_1) + \left(N\chi(2\delta_1) - N^2\chi^2(\delta_1)\right)^2\right]\right].$$

In analogy to the discussion of bunching and superbunching for $\delta_1 = \delta_2$ above, we investigate the position $\delta_1 = \pi$ (see Fig. S2). Considering again even and odd $N$ separately, we obtain

$$g^{(2)}_{n_e,N_{even}}(\pi,-\pi) = g^{(2)}_{n_e,N_{even}}(\pi,\pi)$$
$$g^{(2)}_{n_e,N_{odd}}(\pi,-\pi) = g^{(2)}_{n_e,N_{odd}}(\pi,\pi)\,. \quad (S23)$$

From these expressions it follows that superbunching occurs equivalently for $\delta_1 = \delta_2 = \pi$ and $\delta_1 = -\delta_2 = \pi$.

In case of minimal intensity $\chi(\delta_1) = \chi(2\delta_1) = 0$, i.e., at positions $\tilde{\delta}_1 = a\frac{2\pi}{N}$, where $a = 1,2,\ldots < N/2$, the following form of the correlation function is obtained

$$g^{(2)}_{n_e,N}(\tilde{\delta}_1,-\tilde{\delta}_1) = \frac{(N-1)(4+N^2+2N-6n_e-2Nn_e+2n_e^2)}{(N-2)(n_e-1)n_e}\,. \quad (S24)$$

In principal, for $N \gg n_e$, there is again no upper limit to the correlation function, since $g^{(2)}_{n_e,N}(\tilde{\delta}_1,-\tilde{\delta}_1) \sim N^2/(n_e(n_e-1))$. This behavior is equivalent to $g^{(2)}_{n_e,N_{even}}(\pi,\pi)$ (cf. Eq. (S16)), but valid also for odd $N$. When setting $n_e = 3$, bunching occurs for $N \geq 5$ while superbunching occurs for $N \geq 6$ (cf. Eq. (S17)).

Fig. S2 shows the plotted correlation function $g^{(2)}_{n_e,N}(\delta_1,-\delta_1)$ for three different combinations $(n_e,N) = (2,3),(3,6),(4,8)$. These combinations have been chosen to fulfill the condition for superbunching for minimal number of atoms $N$ at the positions $\tilde{\delta}_1$. Again, it can be observed that the two-photon superradiance decreases with increasing $n_e$, so that higher numbers of $N$ are required to produce correlations comparable to the case $n_e = 2$.

Since for more than two excitations $n_e > 2$ the visibility of $g^{(2)}_{n_e,N}(\delta_1,\delta_2)$ is smaller than one, true antibunching cannot be fulfilled. However, cross correlations smaller than one and therefore nonclassical light can be found for $\delta_1 = -\delta_2 = 0$ in case that $2(n_e - 1) < N$, as has been discussed for $\delta_1 = \delta_2$ (see Eq. (S21)).

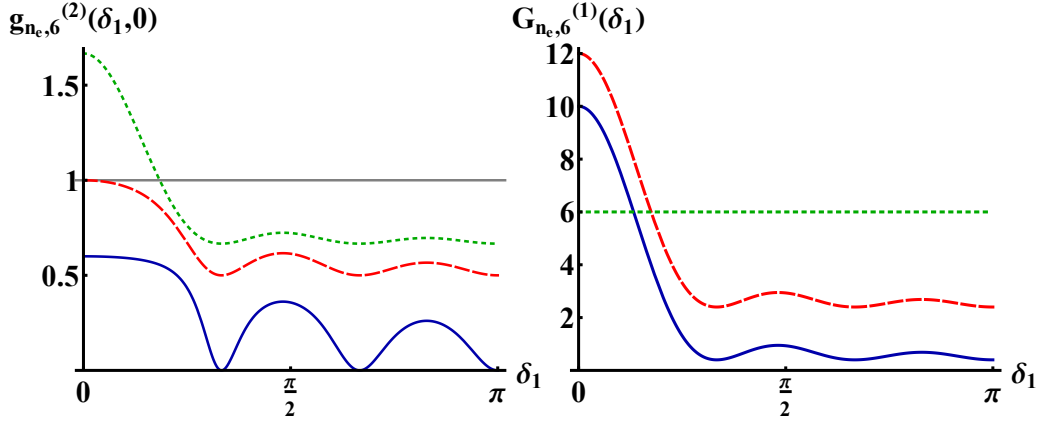

**Figure S3.** Second order correlation function $g^{(2)}_{n_e,N}(\delta_1,0)$ (left) and first order correlation function $G^{(1)}_{n_e,N}(\delta_1)$ (right) for $N=6$ and $n_e = 2$ (solid), $n_e = 4$ (dashed), $n_e = 6$ (dotted). Again, antibunching is observed only for $n_e = 2$, independently of $N$, whereas superbunching can never be obtained (see Eq. (S25)).

Another interesting case is the combination of one detector fixed at position $\delta_2 = 0$ and the second detector moving. Here, the photon cross correlation takes the form (cf. Eq. (S4))

$$g^{(2)}_{n_e,N}(\delta_1,0) = \frac{(n_e-1)(n_e-2)+(n_e-1)(N-n_e+1)N\chi^2(\delta_1)}{n_e(n_e-1)+n_e(N-n_e)N\chi^2(\delta_1)}, \tag{S25}$$

which does not acquire values higher than two so that superbunching can never be obtained (see Fig. S3 for $N=6$). The reason is the following: If $g^{(2)}_{n_e,N}(\delta_1,0) < 1$ the following condition must be fulfilled

$$(2n_e - N - 1)N\chi^2(\delta_1) < 2(n_e-1). \tag{S26}$$

In case of $\chi(\delta_1) = 0$ the relation Eq. (S26) simplifies to

$$0 < 2(n_e - 1), \tag{S27}$$

which is fulfilled for all $n_e \geq 2$. The case $\chi(\delta_1) = 1$ (for $\delta_1 = 0$) already has been investigated in connection with $g^{(2)}_{n_e,N}(0,0)$ and fulfills $g^{(2)}_{n_e,N}(0,0) < 1$ for $2(n_e-1) < N$; for $n_e = N$, $g^{(2)}_{n_e,N}(0,0)$ becomes maximal and reads

$$g^{(2)}_{N,N}(0,0) = 2 - \frac{2}{N} < 2, \tag{S28}$$

what proves the above statement.

As concerns antibunching, according to Eq. (S27), nonclassicality occurs for all $n_e \geq 2$ whereas true antibunching does only occur for $n_e = 2$, as in all other cases the visibility of $g^{(2)}_{n_e,N}(\delta_1,\delta_2)$ remains smaller than one (cf. Eq. (S25)).




\begin{thebibliography}{10}
\bibitem{note(1)}
\bibinfo{title}{Note that in the regime $k d \gg 1$ of widely spaced atoms we
  are investigating, it is simple to obtain $\delta_i = \pi$ or $\delta_i = 2
  \, \pi$, $i = 1, 2$, as these values can be achieved already with observation
  angles $\theta \ll 1$ (see {F}ig.~\ref{fig:detection_scheme}).}
\bibitem{note(2)}
\bibinfo{title}{Note that generally for all detector positions $\delta_{1}$
  fulfilling ${N} \chi^2(\delta_{1})=\chi(2\delta_{1})$ complete antibunching
  is obtained (cf. {E}q.~(\ref{eq:g2ne2d1d1})).}
\end{thebibliography}
\end{document}